\documentclass[11pt]{article}
\usepackage{bbm}

\usepackage{epsfig,endnotes}
\usepackage{multirow}
\usepackage{graphicx}
\usepackage{grffile}

\newcounter{subcopyrightbox@save}
\usepackage[font=bf]{caption}
\usepackage{color, url}
\usepackage{xspace} 

\usepackage{amsmath}

\usepackage{mathrsfs}
\usepackage{amssymb}
\usepackage{amsmath}
\usepackage{amsthm}
\usepackage{epstopdf}
\usepackage{balance}
\usepackage{bm}

\usepackage{thm-restate,thmtools}

\usepackage[ruled,linesnumbered,noend]{algorithm2e}
\SetKwRepeat{Do}{do}{while}
\usepackage{mathtools}

\usepackage{cleveref}
\usepackage{geometry}
\usepackage{changepage}

\usepackage{subfigure}

\usepackage{algorithmic}

\usepackage{makecell}
\usepackage{booktabs}

\newcommand{\RN}[1]{%
  \textup{\uppercase\expandafter{\romannumeral#1}}%
}

\AtBeginDocument{%
  \providecommand\BibTeX{{%
    \normalfont B\kern-0.5em{\scshape i\kern-0.25em b}\kern-0.8em\TeX}}}

\begin{document}

\title{\LARGE{\bf{Reinforcement Learning-based Black-Box Evasion Attacks to Link Prediction in Dynamic Graphs}}}

\author{Houxiang Fan$^{*,1}$, Binghui Wang$^{*,2}$, Pan Zhou$^{3}$, Ang Li$^{2}$, Meng Pang$^{4}$, \\
Zichuan Xu$^{5}$, Cai Fu$^{3}$,  Hai Li$^{2}$, Yiran Chen$^{2}$ \\
  {\footnotesize $^1$School of Electronic Information and Communications, Huazhong University of Science Technology} \\
  {\footnotesize $^2$Department of Electrical and Computer Engineering, Duke University} \\
  {\footnotesize $^3$School of Cyber Science and Engineering, Huazhong University of Science Technology} \\
  {\footnotesize $^4$School of Electrical and Electronic Engineering, Nanyang Technological University} \\
  {\footnotesize $^5$School of Software, Dalian University of Technology} \\
  {\footnotesize $^*$Equal Contribution}
}

   \date{}

   \maketitle

\begin{abstract}
Link prediction in dynamic graphs (LPDG) is an important research problem that has diverse applications such as online recommendations, studies on disease contagion, organizational studies, etc.  
Various LPDG methods based on graph embedding and graph neural networks have been recently proposed and achieved state-of-the-art performance. 
In this paper, we study the vulnerability of LPDG methods and propose the first practical black-box evasion attack. 
Specifically, given a trained LPDG model, our attack aims to perturb the graph structure, without knowing to model parameters, model architecture, etc., such that 
the LPDG model makes as many wrong predicted links as possible.
 We design our attack based on a stochastic policy-based RL algorithm. 
 Moreover, we evaluate our attack on three real-world graph datasets from different application domains. 
 Experimental results show that our attack is both effective and efficient. 
\end{abstract}

\section{Introduction}
\label{sec:intro}

Graphs are often used to describe complex systems such as social networks, biology, social and economic organizations, communication systems, power grid, etc. 
These real-world systems often evolve with time and can be modeled as dynamic graphs, where nodes/entities or links/edges are dynamically added or deleted. 
Links, which represent the interactions
between nodes, are of great importance in the
analysis of dynamic graphs; and one particular important research problem  is called \emph{link prediction in dynamic graphs (LPDG)}. 
Specifically, given historical graph data of a real-world system, LPDG aims to predict its future graph structure so as to better understand the evolution process. 
It is precisely that information in future graphs would be valuable in various applications such as online recommendations, studies on disease contagion, organizational studies, etc.

Various LPDG methods have been proposed in the past decade. 
Conventional methods include feature-based methods~\cite{ibrahim2015link,zhang2017efficient,chiu2018deep,ma2018graph}, generative methods~\cite{sarkar2012nonparametric,yu2018netwalk,zuo2018embedding}, and deep neural networks~\cite{li2014deep,trivedi2017know,chen2019lstm}.
 Recently, graph embedding (GE) methods~\cite{perozzi2014deepwalk,tang2015line,grover2016node2vec} and graph neural networks (GNNs)~\cite{kipf2016semi,velivckovic2017graph,xu2018powerful} have achieved great success in many graph-related tasks (e.g., node classification, link prediction, graph classification, etc.) for static graphs. 
Inspired by this success, many GE/GNN-based LPDG methods ~\cite{li2017attributed,yu2018spatio,nguyen2018continuous,goyal2018dyngem,zhou2018dynamic,pareja2020evolvegcn,goyal2020dyngraph2vec,manessi2020dynamic} have been developed, which extend original GE/GNN methods to the dynamic graphs via introducing a recurrent mechanism that captures the temporal information of dynamic graphs. 
GE/GNN-based LPDG methods have achieved great success in various applications such as forecasting bitcoin user trading~\cite{pareja2020evolvegcn}, traffic prediction~\cite{zhao2019t,yu2018spatio}, predicting loan repayment~\cite{zhou2018dynamic},  forecasting message exchange in communication network~\cite{pareja2020evolvegcn}, etc.
For instance, DyGCN~\cite{manessi2020dynamic} combines LSTM~\cite{hochreiter1997long} and GCN~\cite{kipf2016semi} and achieves 
the state-of-the-art performance.

In this paper, in contrast to designing better LPDG methods, we take the first step to study the vulnerability of LPDG methods. 
In particular, we consider a practical black-box evasion attacks to LPDG--- Given a trained LPDG model, we assume an attacker only knows the predictions via querying the LPDG model, while not knowing the model parameters, model  architecture, etc.  
As to attacker's capability, we consider that the attacker can perturb a certain number of links/edges in the historical graphs, i.e., by adding new edges to or/and deleting existing edges from these graphs.  
Then, the attacker's goal is to learn to perturb the historical graphs such that the LPDG model has as many wrong predicted links as possible in the future graph. 
One possible way to perform such an attack is to formulate the attack as an optimization problem. However, we emphasize that there are two key challenges. First, optimization-based attack requires the attacker knows accurate gradient information related to model parameters, which is difficult to obtain in the black-box setting. 
Second, optimization-based attack for graph structure perturbation is a binary optimization problem, which is NP-hard. 

To address the above challenges, we propose a reinforcement learning (RL)-based attack to LPDG. 
Specifically, we can model perturbing (optimal) edges in a graph as executing (optimal) actions, which is based on solving a policy function in RL. 
Note that solving the policy function only requires LPDG model predictions, and thus our attack does not need to know model parameters.  
Moreover, our attack involves learning a nonlinear mapping from the high-dimensional graph structure space to a low-dimensional representation space. Such a nonlinear mapping can be easily parameterized by a neural network, and naturally fits the RL framework. 
We implement our RL-based attack, specifically to DyGCN, by adopting the soft actor-critic algorithm~\cite{haarnoja2018soft}, which is a stochastic policy-based RL method that has demonstrated a stronger exploration ability and a more stable training. 
We evaluate our attack on three real-world graph datasets from different application domains. 
Our attack is effective. For instance, on Trapping dataset, our attack can reduce the link prediction performance (measured by $F_1$ score) by $37.9\%$ when only $1\%$ of total edges in a graph is perturbed.
Our attack is also efficient. For instance, the running time of our attack is linear to the number of perturbed edges.

\section{Background and Problem Definition}
\label{sec:Preliminary}

\begin{figure*}[tbp]
    \centering 
    \includegraphics[width=0.95\textwidth]{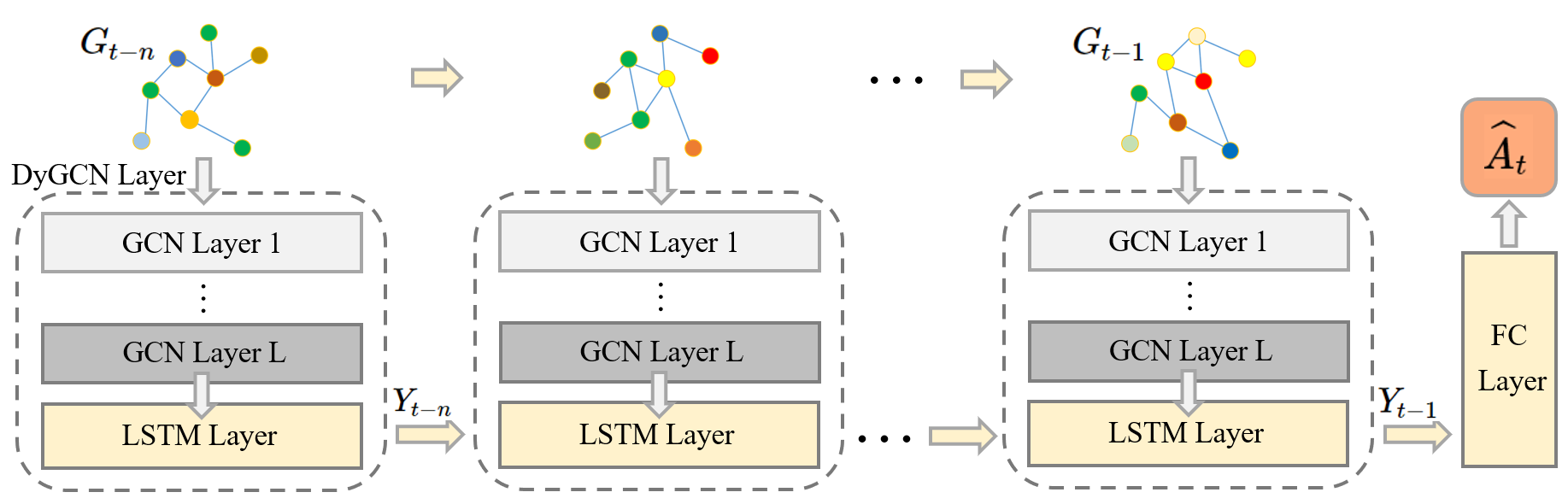} 
    \vspace{-3mm}
    \caption{
    Link prediction in dynamic graphs via DyGCN. We use 
    a single training example $(S_{t-n:t-1}, A_t)$ to illustrate DyGCN. 
    } 
    \label{fig:wolf}
\end{figure*}

\subsection{Link Prediction in Dynamic Graphs}
Given a finite sequence of undirected graphs $\mathcal{G} = \left \{ G_{1},G_{2},...,G_{T} \right \}$, where $G_{t} = (V_t,E_{t}), \forall t \in [1:T]$ denotes a snapshot graph at time step $t$, with $V_t$ the set of nodes and $E_{t}$ the set of edges/links. $|V_t|$ and $|E_t|$ represent the number of nodes and number of edges in graph $G_t$, respectively. W.l.o.g., we assume all graphs in $\mathcal{G}$ have the same node set $V$ in this paper.
We use  $A_{t} \in \left\{0,1\right\}^{|V|\times |V|}$ to  represent the adjacency matrix for $G_t$, i.e., $a_{t}^{(u,v)} = 1$ if there is a link between node $u$ and node $v$ in $G_t$, and $0$ otherwise. \emph{For description convenience, we will use the adjacency matrix and graph interchangeably}.  
For each 
$v \in V$, let $x^{v}_t \in \mathbb{R}^{D}$ be $v$'s feature vector and $X_t = [x_t^{1}; x_t^{2};\cdots; x_t^{|V|}] \in \mathbb{R}^{|V| \times D}$ be the feature matrix of all nodes in graph $G_t$. 

Now, suppose we are given a set of $M_{tr}$ training examples 
$\{(S^{tr}_{[t-n:t-1]}, A^{tr}_{t})\}_{t=n+1}^{n+1+M_{tr}} \subseteq \mathcal{S} \times \mathcal{A}$, where $S^{tr}_{[t-n:t-1]} = \{ A^{tr}_{t-n}, \cdots, A^{tr}_{t-1} \} \in \mathcal{S} $ consists of a sequence of $n$ historical graphs and $A^{tr}_{t} \in \mathcal{A}$ is the future graph to be predicted.  
Then, the goal of link prediction in dynamic graphs is to learn a function $\mathcal{F}_{\Theta}: \mathcal{S} \rightarrow \mathcal{A}$, parameterized by $\Theta$, from the historical graphs in $\mathcal{S}$ to predict links in the future graphs in $\mathcal{A}$. 
In this paper, we focus on DyGCN, as it achieves the state-of-the-art dynamic link prediction performance.

\subsection{Dynamic Graph Convolutional Network (DyGCN)}
DyGCN combines GCN~\cite{kipf2016semi} with LSTM~\cite{hochreiter1997long} to perform link prediction in dynamic graphs. 
GCN was originally developed for a static graph. 
We first briefly review GCN.  
Suppose we are given a  graph $G_{j}$ with adjacency matrix $A_{j}$ and node feature matrix $X_{j}$. GCN stacks multiple (e.g., $L$) graph convolutional layers, and each layer $l$ learns a hidden node feature matrix $H_{j}^{(l)}$. 
Formally, the node feature matrix at $(l+1)$-layer is updated as follows: 
{
\begin{equation*}
    \begin{aligned}
         H_{j}^{(l+1)} =  \sigma(\mathring{A}_{j} H_{j}^{(l)} W^{(l)}),
    \end{aligned}
\end{equation*}
}%
where $\mathring{A}_{j}$ is a normalized version of $A_{j}$, which is defined as $\mathring{A}_{j} = \widetilde{D}_j^{-\frac{1}{2}}\widetilde{A}_{j}\widetilde{D}_j^{-\frac{1}{2}}$, where $\widetilde{A}_{j}=A_{j}+I,\widetilde{D}_j = diag(\sum_{k}\widetilde{A}_{j}^{(i,k)})$; 
$H_{j}^{(0)} = X_{j}$; 
$W^{(l)}$ is the parameter matrix connecting the $l$-th layer and the $(l+1)$-th layer; 
$\sigma(\cdot)$ is a nonlinear activation function, e.g., ReLU . 

In order to perform LPDG, DyGCN introduces multiple GCNs at different time steps and adds an LSTM after each GCN. 
Specifically, at time step $j$, GCN is with an input $A_j$ and the output of GCN (i.e., $H_{j}^{(L)}$) is treated as the input of an LSTM. Moreover, the output of LSTM is denoted as $Y_{j}$.

Now, suppose we have a set of $M_{tr}$ training examples.  
For each training example $(S^{tr}_{[t-n:t-1]}, A^{tr}_t)$, each GCN takes a graph in $S^{tr}_{[t-n:t-1]}$ as an input, e.g., GCN at time step $j$ takes $A^{tr}_{j}$ as an input. 
To predict links in the graph $A^{tr}_t$, 
DyGCN adds a fully connected layer after the \emph{final} output of LSTM at time step $t-1$ (i.e., $Y_{t-1}$), and maps the output to a matrix $Z_{t} \in \mathbb{R}^{|V| \times |V|}$ that has the same size as $A^{tr}_t$. 
Formally, 
{
\begin{equation*}
    Z_{t} = \sigma(fc(Y_{t-1},W_{Z})),
\end{equation*}
}%
where $fc(\cdot)$ is a fully connected layer with parameters $W_{Z}$. As the entries in $Z_{t}$ are usually not binary, DyGCN further performs the following transformation:
{
\begin{equation*}
    \widehat{a}_{t}^{(i,j)} =\left\{\begin{matrix} 
    1, & z_{t}^{(i,j)} > {0.5}; \\ 
    0, & otherwise, \end{matrix}\right.
\end{equation*}
}%
where $\widehat{A}_{t} = \{\widehat{a}_{t}^{(i,j)}\}_{i,j=1}^{|V|}$ is the predicted graph. 

We use the function $\mathcal{F}_{\Theta}$ to encompass all functions involved in DyGCN, where $\Theta$ contains all parameters. Thus, $\widehat{A}_{t} = \mathcal{F}_{\Theta}({S}^{tr}_{[t-n:t-1]})$. 
To train DyGCN, we minimize the loss function $\mathcal{L}$, which is defined as the reconstruction loss between the predicted graph and the ground truth graph for all training examples. 
Specifically, 
{
\begin{align*}
\mathcal{L}(\{\mathcal{F}_{\Theta}({\mathcal{S}^{tr}_{[t-n:t-1]}}),A^{tr}_{t}\}) &= \sum_{t=n+1}^{n+1+M_{tr}} || A^{tr}_{t} - \widehat{A}_{t}||_{F}^2 \odot B_t,
\end{align*}
}%
where the weight matrix $B_t \in \mathbb{R}^{|V| \times |V|}$ is used to mitigate the issue caused by the sparse graph. We set $b_t^{(i,j)} = 1 $ if $a_{t}^{(i,j)}=0$, and $b_t^{(i,j)} = \beta > 1$, otherwise. 
In our experiment, we set $\beta=10$.
Adam~\cite{kingma2014adam} is used to train DyGCN and the learnt model parameters is denoted as $\Theta^{*}$. 
Figure~\ref{fig:wolf} illustrates DyGCN using a single training example.

\subsection{Problem Definition}
We consider black-box evavaion attacks to LPDG and focus on attacking DyGCN in this paper. 
However, we note that our attack is applicable to any 
LPDG algorithm. 
Specifically, given a trained DyGCN model $\mathcal{F}_{\Theta^{*}}$, we assume that the attacker only knows the predictions (i.e., predicted graphs) via querying $\mathcal{F}_{\Theta^{*}}$, while not knowing the model parameters $\Theta^{*}$ or model architecture.  
Moreover, given a testing example $(\mathcal{S}^{te}_{[t-n:t-1]}, A^{te}_t)$ where $\mathcal{S}^{te}_{[t-n:t-1]} = \{A_{t-n}^{te}, \cdots, A_{t-1}^{te}\}$ consists of a 
sequence of $n$ historical graphs, we assume the attacker can perturb a fraction $\mu$ of the $n$ graphs in $\mathcal{S}^{te}_{[t-n:t-1]}$, and each graph is allowed to perturb (e.g., add new edges or remove existing edges) a certain fraction $\delta$ of the total links/edges.  
We denote the perturbed testing example as $\widetilde{\mathcal{S}}_{[t-n:t-1]}^{te} = \{\tilde{A}_{t-n}^{te}, \cdots, \tilde{A}_{t-1}^{te}\}$, where $\tilde{A}_{j}^{te} = {A}_{j}^{te} + \Delta {A}_{j}^{te}$, if ${A}_{j}^{te}$ is perturbed by a binary matrix $\Delta {A}_{j}^{te}$; and $\tilde{A}_{j}^{te} = {A}_{j}^{te}$, if not. 
For simplicity, we consider recent historical graphs, i.e., 
$[{A}_{t-\mu n}^{te}:{A}_{t-1}^{te}]$, are perturbed. 
Then, the attacker's goal is learn to perturb the graphs,  such that the link prediction performance on $\widetilde{\mathcal{S}}_{[t-n:t-1]}^{te}$ evaluated by $\mathcal{F}_{\Theta^{*}}$ is maximally decreased. 
Formally, the attacker's objective function is defined as follows: 
{
\begin{align}
        \label{obj:attack}
        & \max_{\{\Delta {A}^{te}_{j}\}}  \, 
        \textrm{err} = \mathcal{I}({A}_{t}^{te} \neq \mathcal{F}_{\Theta^{*}}(\widetilde{\mathcal{S}}^{te}_{[t-n:t-1]})), 
        \quad \textrm{s.t. } \,
        | \Delta {{A}^{te}_{j}} | \leq \delta \cdot |E_j|, \quad \forall j \in [t-\mu n:t-1],
\end{align}
}%
where $\mathcal{I}(A \neq B)$ is a function that counts the total number of unequal elements between $A$ and $B$ at the same position. 

Directly solving Equation~\ref{obj:attack} to achieve the attacker's goal is challenging, as it is a binary optimization problem that is NP-hard. 
In the next section, we will introduce our RL-based attack against DyGCN.

\section{Our Proposed RL-based Attack}
\label{sec:Attack Algorithms}

\begin{figure*}[htbp]
    \centering 
    \includegraphics[width=0.95\textwidth]{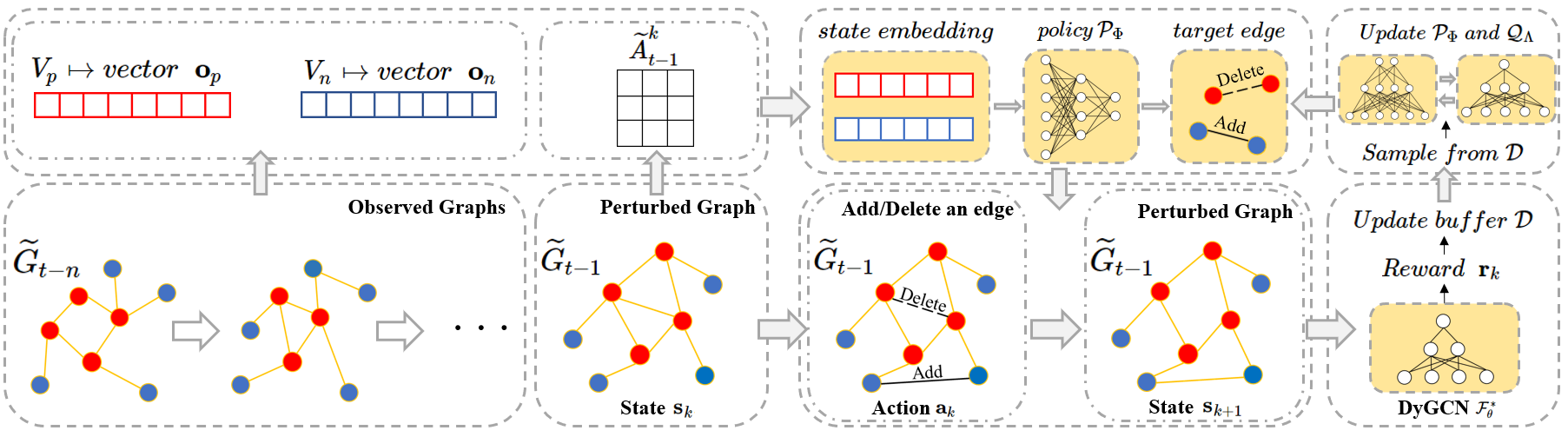}
    \vspace{-3mm}
    \caption{Overview of our RL-based black-box attack to DyGCN. We take attacking a snapshot graph $G_{t-1}$ as an example.   
    } 
    \label{fig:attack} 
\end{figure*}

In this section, we propose to use RL to perform black-box evasion attacks to DyGCN.   
Our RL-based attack is based on soft actor-critic (SAC)~\cite{haarnoja2018soft}, which is a stochastic policy-based RL method and achieves state-of-the-art performance.
Comparing with optimization-based attacks, RL-based attacks have two major advantages. 
First, perturbing (optimal) edges in a graph (i.e., add new edges or delete existing edges) can be naturally modeled by executing (optimal) actions based on solving a policy function in RL, while solving which is NP-hard using optimization methods. 
Second, our attack involves learning a nonlinear mapping from the  high-dimensional graph structure space to a low-dimensional representation space. Such a nonlinear mapping can be easily parameterized by a neural network and fits the RL framework. 

The core idea of our attack is as follows: First, the attacker observes the current state associated with a graph. Based on the current state, the attacker executes an optimal action (i.e., add new edges and delete existing edges) by solving a policy function parameterized by a policy network; and obtaining a reward, which relates to 
our objective function in Equation~\ref{obj:attack}, by solving a soft $Q$-function parameterized by a $Q$-network. 
Then, the attacker obtains the next state and saves the current and next states, actions, and rewards as a trajectory in a replay buffer. 
Finally, the attacker samples the trajectories from the buffer and adopts the SAC algorithm to train our attack. Figure~\ref{fig:attack} shows our proposed RL-based attack to DyGCN. 
The used notations and their descriptions are listed in Table~\ref{tbl:nota}. 

\subsection{The Attack Environment}
A RL method consists of states, actions, rewards, and terminal condition. 
We first define the attack environment from the attacker's perspective, which involves how to represent states, execute actions, define rewards, and determine the terminal condition.   
We take attacking a single graph $G_{t-1}$ ($A_{t-1}$) in an  example $S_{[t-n:t-1]}=\{A_{t-n},\cdots, A_{t-1}\}$ as an instance to introduce our attack. 

\subsubsection{States.} 
We use $\textbf{s}_{k}$ to denote the state of the intermediate perturbed graph $\widetilde{A}_{t-1}^k$ at the attack step $k$. Thus $\textbf{s}_{0}$ is the state of the clean graph $A_{t-1}$ and $\widetilde{A}_{t-1}^0=A_{t-1}$. 
Moreover, we also use ${\textbf{s}}_{k}$  to indicate a low-dimensional embedding of the high-dimensional perturbed graph $\widetilde{A}^k_{t-1}$. 
${\textbf{s}}_{k}$ aims to capture the latent structural information in the perturbed graph $\widetilde{A}_{t-1}^k$. 
 and it is learnt via the following three steps:
First, we observe all clean graphs in the example other than ${A}^k_{t-1}$ (i.e., from $A_{t-n}$ to $A_{t-2}$ in this case), and select a fraction $\rho$ ($<0.5$) of nodes $V_{p} \subseteq V$ with the largest averaged degrees (called \emph{popular nodes}) as well as the same fraction of nodes $V_{n} \subseteq V $ with the smallest averaged degrees (called \emph{neglected nodes})  
among those observed graphs (i.e., $|V_{p}| = |V_{n}| = \rho |V|$). Our intuition is that nodes with the largest/smallest averaged degrees can maintain the primary information in a graph. 
Second, we encode the two node sets $V_{p}$ and $V_{n}$ using two one-hot vectors $\textbf{o}_{p},\textbf{o}_{n} \in \mathbb{R}^{|V|}$. Specifically, $o_{p,u} = 1$ if $u \in V_{p}$ and $o_{p,u} = 0$ if $u \notin V_{p}$. 
Similarly, $o_{n,u} = 1$ if $u \in V_{n}$ and $o_{n,u} = 0$ if $u \notin V_{n}$. 
Third, we define the representation for ${\textbf{s}}_{k}$ as follows: 
{
\begin{equation}
    \begin{aligned}
        & {\textbf{s}}_{k} = \left[ {\textbf{s}}_{k}^{(1)}, {\textbf{s}}_{k}^{(2)} \right], \\
        & {\textbf{s}}_{k}^{(1)} = g(\widetilde{A}_{t-1}^{k} \cdot ((1- \gamma) \textbf{o}_{p} + {\gamma}/{|V|})), \\
        & {\textbf{s}}_{k}^{(2)} = g(\widetilde{A}_{t-1}^{k} \cdot ((1- \gamma) \textbf{o}_{n} + {\gamma}/{|V|})),
        \label{eqn:state}
    \end{aligned}
\end{equation}
}%
where $g(a)$ is a function $g: \mathbb{R}^{|V|} \rightarrow \mathbb{R}^{\rho |V|}$, which first selects $\rho |V|$ entries from the $|V|$-dimensional vector $a$ with indexes corresponding to $V_{p}$ (or $V_{n}$) to form a new $\rho |V|$-dimensional vector; and then applies a nonlinear activation function (e.g., ReLU in our paper) to the new vector to obtain the $\rho |V|$-dimensional embedding for the perturbed graph $\tilde{A}^k_{t-1}$. 
Note that we introduce a smooth regularization term $\gamma/|V|$ in order to stabilize the training and reduce overfitting. 
By default, we set $\gamma$ to be 0.2.
With Equation~\ref{eqn:state}, each entry value in ${\textbf{s}}_{k}$ indicates the importance of the corresponding node in $V_{p}$ or $V_{n}$. For the implementation purpose, we also store the mapping between the vector index $i \in [1:V_{p}]$ or $i \in [1:V_{n}]$ and the real node index $j \in V$ in the graph.

\begin{table}[tbp]
\centering
    \caption{\textbf{Notations and descriptions.}}
    \scalebox{0.9}{
    \begin{tabular}{ll}
        \toprule
       Notation  &  Description\\
       \midrule
       $\mathcal{S}^{tr}/\mathcal{S}^{val}/\mathcal{S}^{te}$ & Training/Validation/testing set \\
       $M_{tr}/M_{val}/M_{te}$ & No. of Training/Validation/testing examples \\
       $\mathcal{F}_{\Theta^{*}}$ & Learnt DyGCN model \\
       $G_j/{A}_{j}$ & Ground truth graph at time step $t$  \\
       $\widehat{G}_{j}/\widehat{A}_{j}$ & Estimated graph at time step $t$ \\
       $\widetilde{G}_{j}/\widetilde{A}_{j}$ & Perturbed graph at time step $j$ \\
       $V_{p}/V_{n}$ & Popular/Neglected node set\\
       $n$ & No. of graphs in a sequence \\
       $N$ & No. of nodes in each graph \\
       $L$ & No. of layers in GCN \\
       $\mathcal{P}_{\Phi}$ & Policy network\\
       $\mathcal{Q}_{\Lambda}$ & Soft $\mathcal{Q}$-network\\
       $\mathcal{D}$ & Replay buffer\\
       $\alpha$ & Temperature hyperparameter \\
       $\mu$ & Fraction of perturbed graphs in an example \\
       $\rho$ & Fraction of selected popular/neglected nodes \\
       $\delta$ & Fraction of total edges perturbed in a graph\\
       $T$  & No. of training episodes \\
       \bottomrule
    \end{tabular}
    }
    \label{tbl:nota}
    \vspace{-4mm}
\end{table}

\subsubsection{Action.} 
We consider that the attacker only changes the link status among 
$V_{p}$ and $V_{n}$. 
Our motivation is based on the assumption that: when deleting edges among popular nodes or/and adding edges among neglected nodes, the graph structure can be largely changed, and thus the  DyGCN model would make as many wrong predicted links as possible when evaluated on the perturbed graph. 
Note that such a consideration can also reduce the computational complexity, as the attacker does not need to search the entire graph structure. 
Moreover, to ensure that certain graph property does not change significantly after the perturbation, we require the total number of edges in the graph keeps the same in each attack step. 
Particularly, we consider that an attacker executes an action by adding a new edge between a pair of nodes as well as deleting an existing edge between another pair of nodes. 
Then, the purpose of the action is to search two pairs of nodes, such that when executing the action, 
DyGCN's performance decreases as much as possible based on the current state. 
Specifically, we denote $\textbf{a}_{k}$ as the action at the attack step $k$. 
Then, we design a policy to generate actions based on 
$\textbf{s}_{k}$.
Formally, we define
\begin{equation}
    \textbf{a}_{k} \sim \mathcal{P}_{\Phi} (\textbf{a}_{k} | \textbf{s}_{k}),
    \label{obj:action}
\end{equation}
where $\mathcal{P}_{\Phi}$ is a policy network parameterized by $\Phi$. 
Equation~\ref{obj:action} attempts to enforce that, when  nodes in $V_p$ (or $V_n$) are relative more important at state $\textbf{s}_{k}$, then they are more likely to be selected as the action $\textbf{a}_{k}$.  
In this paper, we define our policy network as a composition of a 3-layer MLP network and a truncated normal distribution (TruncNorm) within an interval $[1, \rho |V|]$.
Specifically, the output of MLP is the mean and log-deviation of a TruncNorm;  
and we then perform the action $\textbf{a}_{k}$ by sampling two pairs of nodes from the TruncNorm (with rounding). 
Moreover, we note that the sampling process can be naturally divided into two separate steps: 
we first perform an action $\textbf{a}^{(1)}_{k}$ to sample a pair of nodes from $V_{p}$ and then perform another action $\textbf{a}^{(2)}_{k}$ to sample a pair of nodes from $V_{n}$. Formally, 
{
\begin{equation*}
\begin{aligned}
    & \textbf{a}_{k} = [\textbf{a}^{(1)}_{k}, \textbf{a}^{(2)}_{k}] \in \mathbb{R}^{2 \times 2}, \quad \Phi = [\Phi_1, \Phi_2] \in \mathbb{R}^{\rho |V| \times 2}, \\
    & \textbf{a}_{k}^{(1)}  \sim \textrm{TruncNorm}(\textbf{a}_{k}^{(1)} | \mu_1, \textrm{diag}(\sigma_1^2), 1, \rho |V|]).\textrm{round()}, \\
    & \mu_1 = \textrm{MLP}_{\Phi_1} (\textbf{s}_{k}^{(1)})_{1:2} \in \mathbb{R}^2, \, \log \sigma_1 = \textrm{MLP}_{\Phi_1} (\textbf{s}_{k}^{(1)})_{3:4} \in \mathbb{R}^2, \\
    & \textbf{a}_{k}^{(2)} \sim 
    \textrm{TruncNorm}(\textbf{a}_{k}^{(2)} | \mu_2, \textrm{diag}(\sigma_2^2), 1, \rho |V|).\textrm{round()}, \\
    & \mu_2 = \textrm{MLP}_{\Phi_2} (\textbf{s}_{k}^{(2)})_{1:2} \in \mathbb{R}^2, \, \log \sigma_2 = \textrm{MLP}_{\Phi_2} (\textbf{s}_{k}^{(2)})_{3:4} \in \mathbb{R}^2. 
\end{aligned}
\end{equation*}
}%
With $\textbf{a}_{k}$, we can map each of its value back to the real node index via our stored mapping. For simplicity, we still use $\textbf{a}_{k}$ to indicate the real node indexes.  
Moreover, when the attacker's action is to add an existing edge or delete a nonexistent edge, we will move to the next step. Details of training the policy network $\mathcal{P}_{\Phi}$ is illustrated in the next subsection. 

\subsubsection{Rewards.}
Given the state $\textbf{s}_{k}$ and action $\textbf{a}_{k}$, the attacker obtains a new state  $\textbf{s}_{k+1}$.
Naturally, if the link prediction performance of $\mathcal{F}_{\Theta^{*}}$ at state $\textbf{s}_{k+1}$ is worse than that at state $\textbf{s}_{k}$, then a positive reward will be given, otherwise a negative reward. 
Based on this motivation, we design the reward function as follows:
{
\begin{equation}
    \begin{aligned}
        \textbf{r}_{k}(\textbf{s}_{k},\textbf{a}_{k}) &= \left\{\begin{matrix} f^{k} - f^{k+1}  &  \qquad \textrm{if } \textrm{err}^{k+1} > \textrm{err}^{k}; \\
        -\mu \cdot n & \textrm{otherwise},  \end{matrix}\right. 
    \label{obj:reward}
    \end{aligned}
\end{equation}
}%
where $\textrm{err}^{k} = \mathcal{I}({A}_{t}^k \neq \mathcal{F}_{\Theta^{*}}(\widetilde{\mathcal{S}}^k_{[t-n:t-1]}))$ corresponds to our objective function in Equation~\ref{obj:attack} at the $k$-th attack step. 
$f^k$ is a function to measure the effectiveness of $\mathcal{F}_{\Theta^{*}}$ on the perturbed graphs for link prediction at state $\textbf{s}_{k}$. If $\textrm{err}^{k+1} > \textrm{err}^{k}$, which means our attack generates more wrong predicted links at the $k+1$-th attack step, then the gap between $f^{k}$ and $f^{k+1}$ is positive. Furthermore, if the positive gap is larger, then our attack is more effective. Thus, we can use this positive gap as a positive reward $\textbf{r}_k$. 
Otherwise, if our attack performs worse at the attack step $k+1$, 
then we set a large negative reward to penalize this step. 

To obtain a optimal expected reward and guide the policy network $\mathcal{P}_{\Phi}$, we also need to solve a $Q$-function, which is 
parameterized by a soft $\mathcal{Q}$-network $\mathcal{Q}_{\Lambda}$. 
In our paper, soft $\mathcal{Q}$-network uses the same MLP as the policy network.
Specifically, $\mathcal{Q}$-network takes a (state, action) pair as an input and outputs a score that indicates the effectiveness of the action. Its associated Bellman equation is defined as:
{
\begin{equation}
    \mathcal{Q}_{\Lambda}(\textbf{s}_{k},\textbf{a}_{k}) := \textbf{r}_{k}(\textbf{s}_{k},\textbf{a}_{k}) + \lambda\mathbb{E}_{\textbf{s}_{k+1}}[\mathcal{V}(\textbf{s}_{k+1})],
\end{equation}
}%
where
{
\begin{equation}
    \mathcal{V}(\textbf{s}_{k}) = \mathbb{E}_{\textbf{a}_{k}\sim \mathcal{P}_{\Phi}} [Q_{\Lambda}(\textbf{s}_{k},\textbf{a}_{k}) - \alpha \log \mathcal{P}_{\Phi}(\textbf{a}_{k}|\textbf{s}_{k})]
\end{equation}
}%
is the state value function that denotes the expected reward obtained by the attacker at state $\textbf{s}_{k}$. The discount factor $\lambda = 0.99$ is to limit the sum of the expected rewards.
Details of training the soft $Q$-network is shown in the next subsection.

\subsubsection{Terminal.}
As required for the attacker, the number of modified edges for each perturbed graph in the sequence is limited to $\delta$. During the training process, the steps for each training episode is finite and fixed.  
In each attack step, the attacker agent adds a new edge as well as deletes an existing edge, therefore there are at most $\delta/2$ steps per episode.

\subsection{Maximum Entropy Reinforcement Learning }
We use the soft actor critic (SAC) algorithm~\cite{haarnoja2018soft}, a method of maximum entropy reinforcement learning (MERL), to train our attack model. This algorithm maximizes the entropy of the policy as well as the expected reward. 
Compared with other RL algorithms, SAC has a stronger exploration ability and a more stable training. 
We first define the loss function of the policy network $\mathcal{P}_{\Phi}$ and the soft $\mathcal{Q}$-network $\mathcal{Q}_{\Lambda}$;
and then show how to train them. 

\subsubsection{Policy Network.}
Its loss function is defined as
{
\begin{align}
\mathcal{L}_{\mathcal{P}_{\Phi}} = 
\mathbb{E}_{\textbf{s}_{k}\sim\mathcal{D}} \left[ \mathbb{E}_ {\textbf{a}_{k} \sim \mathcal{P}_{\Phi}} \left[ \alpha \log (\mathcal{P}_{\Phi}(\textbf{a}_{k}|\textbf{s}_{k})) - \mathcal{Q}_{\Lambda}(\textbf{s}_{k}, \textbf{a}_{k}) \right] \right],
\label{loss:P}
\end{align}
}%
where $\mathcal{D}$ is a replay buffer which collects a set of previous states, actions, and rewards. $\alpha$ is a temperature parameter that controls the stochasticity of the optimal policy. In practice, $\alpha$ is an important parameter that 
is learnt as follows:   
{
\begin{equation}
    \mathcal{L}_{\alpha} = \mathbb{E}_{\textbf{a}_{k}\sim \mathcal{P}_{\Phi}}[-\alpha  \textup{log}\mathcal{P}_{\Phi}(\textbf{a}_{k}|\textbf{s}_{k}) - \alpha \mathcal{H}_{0}],
\end{equation}
}%
where $\mathcal{H}_{0}$ is a predefined entropy threshold (e.g., $\mathcal{H}_{0}=-2$ in our paper).  
Minimizing loss function in Equation~\ref{loss:P} can make the policy $\mathcal{P}_{\Phi}$ select multiple attractive actions whose probabilities are similar, instead of focusing on a single determinate action.

Furthermore, to estimate the density of the  $\mathcal{Q}$-function with a lower variance estimator, we usually apply a reparameterization trick to reparameterize the policy using a neural network transformation $f_{\Phi}(\epsilon_{k};\textbf{s}_{k})$, 
where $\epsilon_{k}$ is an input noise, sampled from, e.g., a normal Gaussian distribution. Then, we can rewrite the loss in Equation~\ref{loss:P} as 
{
\begin{equation}
\begin{aligned}
    \mathcal{L}_{\mathcal{P}_{\Phi}} &= \mathbb{E}_{\textbf{s}_{k}\sim\mathcal{D},\epsilon_{k}\sim\mathcal{|V|}}[\alpha \cdot \textup{log}\mathcal{P}_{\Phi}(f_{\Phi}(\epsilon_{t};\textbf{s}_{k})|\textbf{s}_{k}) -  \mathcal{Q}_{\Lambda}(\textbf{s}_{k},f_{\Phi}(\epsilon_{k};\textbf{s}_{k}))].  
\end{aligned}
\end{equation}
}%

\begin{algorithm}[t]
\caption{Train our attack model on a validation set.} 
\label{alg:Framwork} 
\begin{algorithmic}[1] 
\REQUIRE ~~\\ 
A set of $M_{val}$ validation examples $(\mathcal{S}^{val}, \mathcal{A}^{val}) = \{({S}^{val}_{[t-n:t-1]}, A^{val}_t)\}_{t=n+1}^{n+1+M_{val}}$. 
Trained DyGCN model: $\mathcal{F}_{\Theta}^*$.
Hyperparameters: $T, \mu, \rho, \delta, n$.

\ENSURE ~~\\ 
Our trained attack model $\Theta_{att} = \{\Phi^{*}, \alpha^{*}, \Lambda^{*}\}$.

\STATE Initialize the model parameters $\Phi$, $\alpha$, $\Lambda$;

\STATE Initialize the replay buffer $ \mathcal{D}$;

\FOR{each ${S}^{val}_{[t-n:t-1]} \in \mathcal{S}^{val}$}
\STATE Select $\mu$ fraction of $n$ graphs (denoted as  ${S}^{val}_\mu$) from ${S}^{val}_{[t-n:t-1]}$ to be attacked;  

\STATE Select ${V_{p}}$ and ${V_{n}}$ using the other $(1-\mu) n$ graphs;

\FOR{each graph $A_{j} \in \mathcal{S}^{val}_\mu$}
\STATE Initialize the perturbed graph $\widetilde{A}_{j}^{0}\leftarrow A_{j}$;

\WHILE{episode $ \text{epi} \leq {T}$}
 
\WHILE{attack step $k \leq \delta/2$}

\STATE Learn $\textbf{s}_k$ based on $\widetilde{A}_{j}^{k}$ via Eq. (2);

\STATE Sample an action $\textbf{a}_{k}$ at state $\textbf{s}_{k}$ via Eq. (3);

\STATE Obtain reward $\textbf{r}_{k}$ via Eq. (4) and  $\mathcal{F}_{\Theta}^*$;

\STATE Update state $\textbf{s}_{k+1} = \{\textbf{s}_k, \textbf{a}_k \}$;

\STATE Update perturbed graph $\widetilde{A}_{j}^{k+1} \leftarrow \widetilde{A}_{j}^{k} \bigcup \textbf{a}_k $;

\STATE Update buffer $\mathcal{D}$ $\leftarrow$ $\mathcal{D}$  $\cup$ 
$\left\{\textbf{s}_{k},\textbf{a}_{k},\textbf{r}_{k},\textbf{s}_{k+1}\right\}$;

\STATE Sample a batch of trajectories from $\mathcal{D}$;

\STATE Update $\Phi$, $\alpha$, $\Lambda$ according to Eq.(7)-Eq.(10).
\ENDWHILE
\ENDWHILE
\ENDFOR
\ENDFOR
\RETURN Attack model parameters $ \{\Phi^{*}, \alpha^{*}, \Lambda^{*}\}$. 
\end{algorithmic}
\end{algorithm}

\subsubsection{Soft $\mathcal{Q}$-Network.}
Its loss function is defined as:
{
\begin{equation}
    \begin{aligned}
        \mathcal{L}_{\mathcal{Q}_{\Lambda}} &= \mathbb{E}_{(\textbf{s}_{k},\textbf{a}_{k}) \sim \mathcal{D}}
        [ \frac{1}{2} (\mathcal{Q}_{\Lambda}(\textbf{s}_{k},\textbf{a}_{k})-\textbf{r}_{k}(\textbf{s}_{k},\textbf{a}_{k}) -\lambda (\mathcal{Q}_{\Lambda }(\textbf{s}_{k+1},\textbf{a}_{k+1}) +
        \alpha \log(\mathcal{P}_{\Phi}(\textbf{a}_{k+1}|\textbf{s}_{k+1}))))^{2}], 
    \end{aligned}
\end{equation}
}%
where $\textbf{s}_{k}$ and $\textbf{a}_{k}$ are sampled from the relay buffer $\mathcal{D}$, and $\textbf{a}_{k+1}$ is sampled from the policy during the training process.

\subsubsection{Attack model training.} 
After defining the loss function of the parameterized policy network, parameterized soft $\mathcal{Q}$-network and temperature parameter, we can now train our RL-based attack model. 
Suppose we are given a trained DyGCN model $\mathcal{F}_{\Theta^*}$, a validation set, and a testing set.  
We use the validation set and the Adam algorithm to train our attack against the DyGCN model.
Specifically, we first update the parameters $\Phi$ in the policy network $\mathcal{P}_{\Phi}$, which guides the policy improvement; Then, we update the temperature parameter $\alpha$ based on updated parameters $\Phi$.  
Third, we update the parameter $\Lambda$ in the soft $\mathcal{Q}$-network, which evaluates the effectiveness of the policy. 
We iteratively and alternatively update these parameters until reaching the terminal condition. 
Finally, we can evaluate our trained attack model on a testing set. 
The training process and evaluation process of our attack are detailed in Algorithm 1 and Algorithm 2, respectively.

\begin{algorithm}[t] 
\caption{Evaluate our attack model on a testing set.} 
\label{alg:Framwork} 
\begin{algorithmic}[1] 
\REQUIRE ~~\\ 
A set of $M_{te}$ testing examples $(\mathcal{S}^{{te}}, \mathcal{A}^{{te}}) = \{({S}^{{te}}_{[t-n:t-1]}, A^{{te}}_t)\}_{t=n+1}^{n+1+M_{{te}}}$. 
Trained DyGCN model: $\mathcal{F}_{\Theta}^*$.
Trained attack model $\Theta_{att}$.
Parameters: $\mu$, $\rho$, $\delta$, $n$.
\ENSURE ~~\\ 
Perturbed testing examples $\widetilde{\mathcal{S}}^{te}$ and score  $f(\widetilde{\mathcal{S}}^{te})$.

\FOR{each $\mathcal{S}_{[t-n:t-1]}^{te} \in \mathcal{S}^{te}$}

\STATE Select $\mu$ fraction of $n$ graphs (denoted as $\mathcal{S}_{\mu}^{te})$ from $\mathcal{S}_{[t-n:t-1]}^{te}$ to be attacked;

\STATE Select $V_{p}$ and $V_{n}$ using the other $(1-\mu)n$ graphs;

\FOR{each $A_{j} \in \mathcal{S}_{\mu}^{te}$}
\STATE Initialize the perturbed graph $\widetilde{A}_{j}^{0}\leftarrow A_{j}$;

\WHILE{attack step $k \leq \delta/2$}

\STATE Obtain state $\textbf{s}_k$ based on Eq. (7) and action $\textbf{a}_{k}$ using the attack model  $\Theta_{att}$ and $\mathcal{F}_{\Theta}^*$;

\STATE Update perturbed graph $\widetilde{A}_{j}^{k+1} \leftarrow \widetilde{A}_{j}^{k} \bigcup \textbf{a}_{k}$;

\ENDWHILE
\ENDFOR
\ENDFOR
\RETURN $\widetilde{\mathcal{S}}^{te}$, $f(\widetilde{\mathcal{S}}^{te})$ 
\end{algorithmic}
\end{algorithm}

\section{Evaluation}
\label{sec:experiments}

\begin{table}[tbp]\renewcommand{\arraystretch}{1.3}
    \centering
    \caption{\textbf{Dataset statistics.}
    \vspace{-3mm}
  }  

    \begin{tabular}{c|ccccc}
        \toprule
        Dataset & Nodes & Edges & Graphs \\
        \midrule
        Haggle & 274 & 28.2k & 30 \\
        Hypertext & 113 & 20.8k & 30 \\
        Trapping & 1.5k & 4.6k & 30 \\
        \bottomrule
    \end{tabular}
\end{table}

\subsection{Experimental Setup}

\subsubsection{Dataset description.} 
We evaluate our RL-based black-box evasion attack to DyGCN, a state-of-the-art LPDG method, on three graph datasets, i.e., Haggle, Hypertext, and Trapping, from three different domains. We split each dataset into 30 graphs in a chronological order.
\begin{itemize}
    \item \textit{Haggle}. This is a social network available at KONECT\footnote{http://konect.uni-koblenz.de/networks/contact}, representing the connection between users measured by wireless devices. A node represents a user and an link between two person indicates a contact between them. The dataset was collected within 5 days. 
    
    \item \textit{Hypertext}. This is a face-to-face contact network of ~100 attendees to the ACM Hypertext 2009 conference held in Turin, Italy over three days from June 29 to July 1, 2009. The network is provided by SocioPatterns$\footnote{http://www.sociopatterns.org/datasets}$. In the network, a node represents a conference attendee, and an edge represents a face-to-face contact between two attendees that was active for at least 20 seconds. 
    
    \item \textit{Trapping}. This is a real-world animal interaction network from Network Data Repository\footnote{http://networkrepository.com/mammalia-voles-rob-trapping.php}. The animal interaction data were from published studies of wild, captive, and domesticated animals. Each node represents an animal, a mammal, or a voles. An edge was added into the network whenever two voles were caught in at least one common trap over the primary trapping sessions. Each network in the dataset has a 12-hour time span. 

\end{itemize}

\begin{figure*}[t]
    \centering
    \subfigure[Haggle]{
    \includegraphics[width=0.3\textwidth]{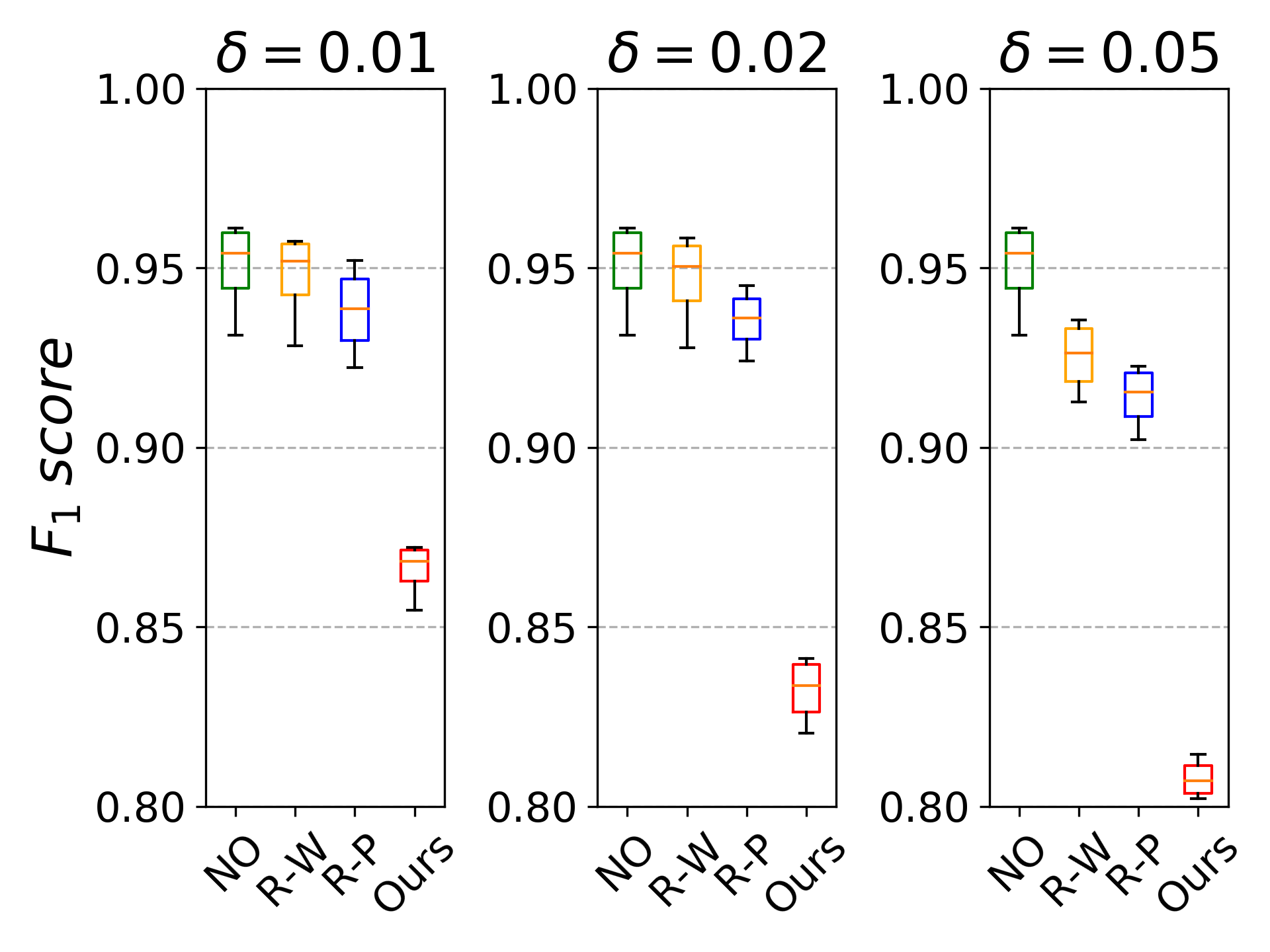}
    }
    \subfigure[Hypertext]{
    \includegraphics[width=0.3\textwidth]{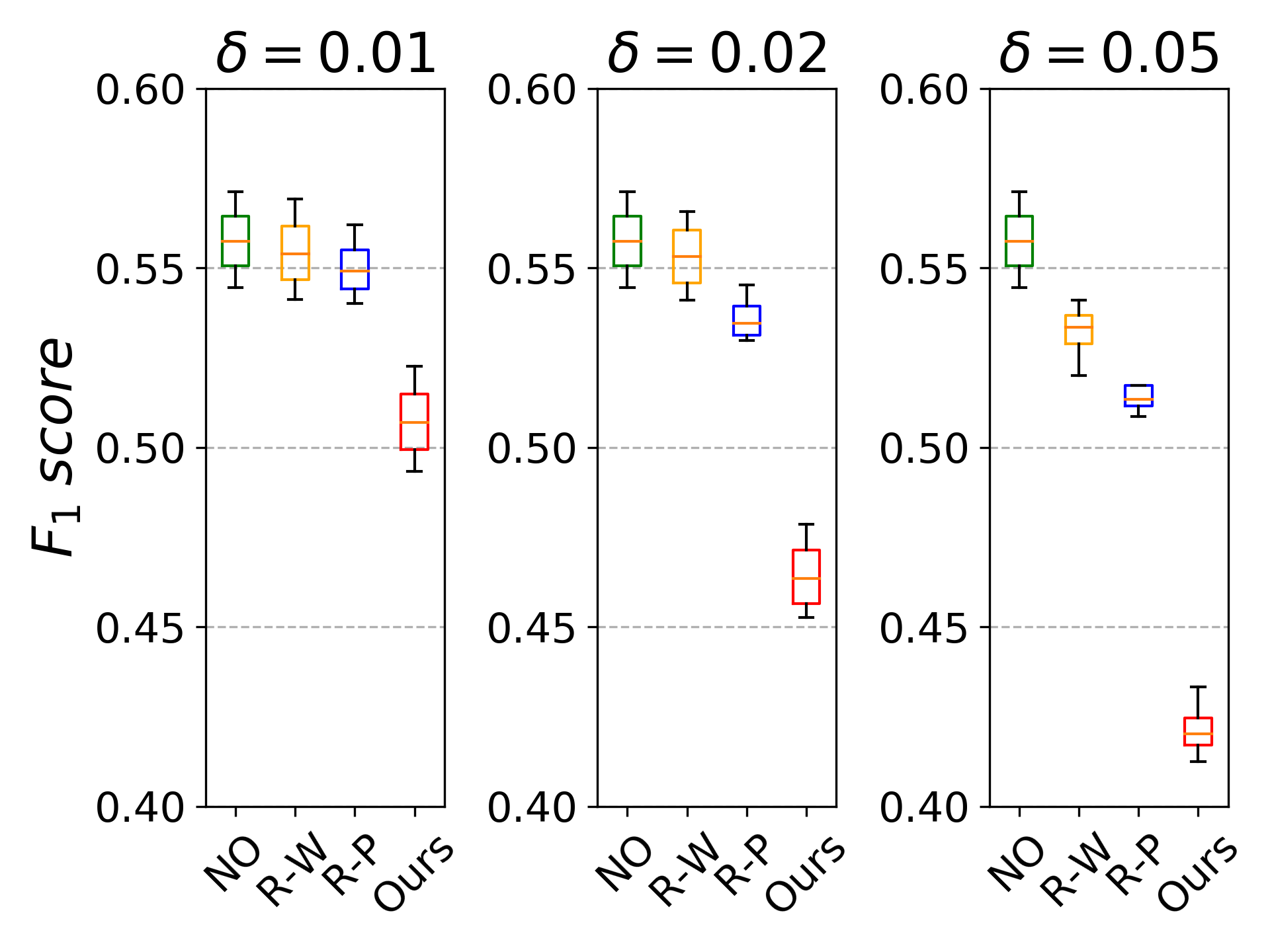}
    }
    \subfigure[Trapping]{
    \includegraphics[width=0.3\textwidth]{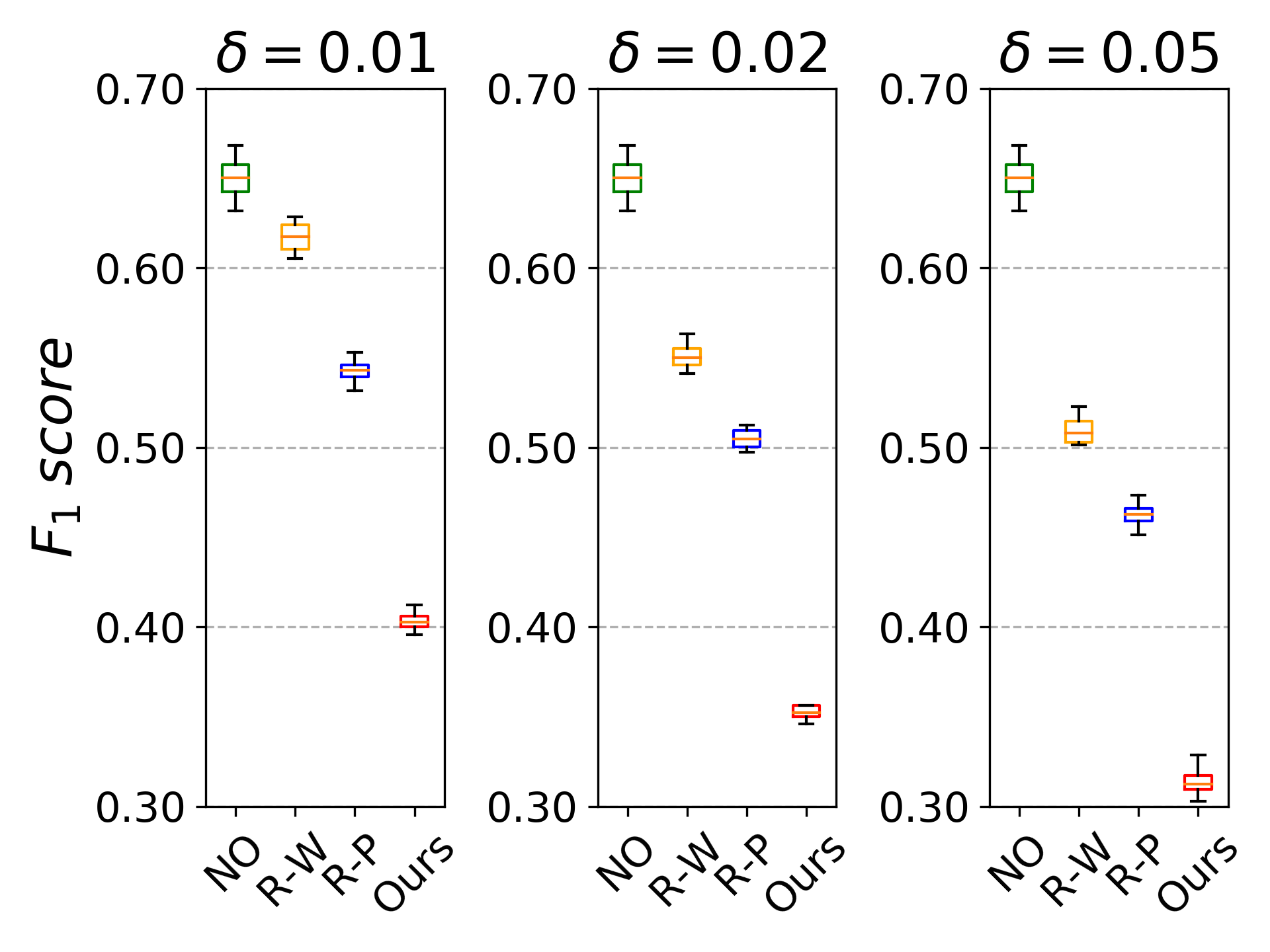}
    }
    \vspace{-3mm}
    \caption{Impact of $\delta$ $(\rho = 0.30,\delta = 0.02)$. 
    } 
    \label{fig:impact_delta}
\end{figure*}

\begin{figure*}[t]
    \centering
    \subfigure[Haggle]{
    \includegraphics[width=0.3\textwidth]{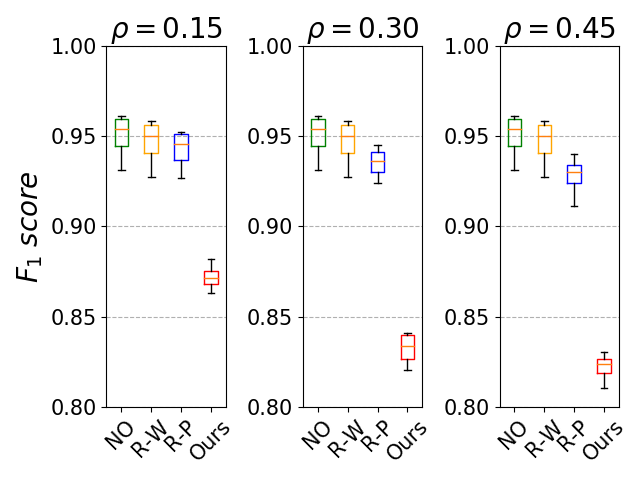}
    }
    \subfigure[Hypertext]{
    \includegraphics[width=0.3\textwidth]{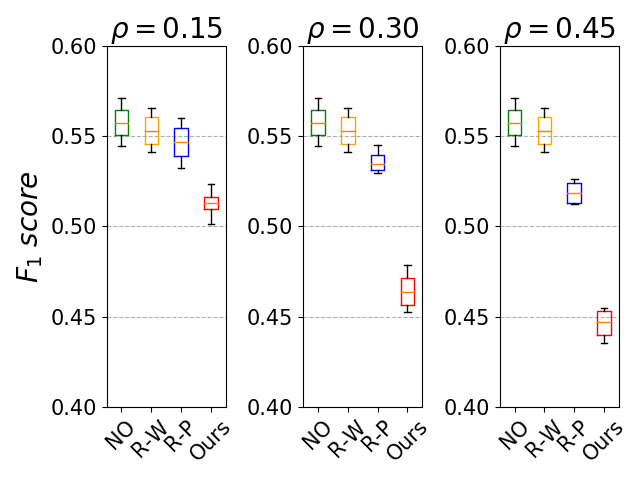}
    }
    \subfigure[Trapping]{
    \includegraphics[width=0.3\textwidth]{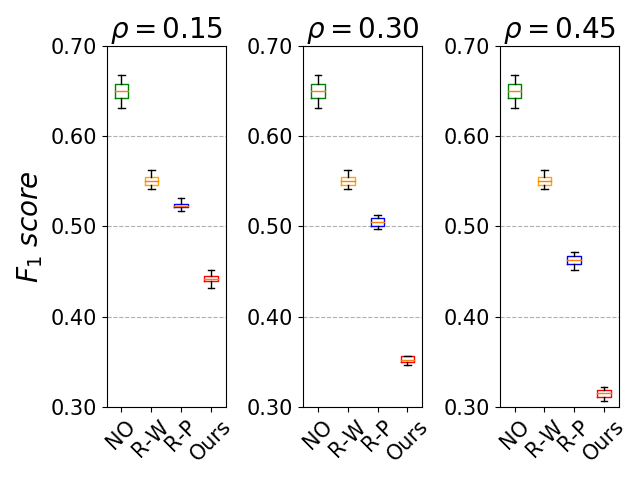}
    }
    \vspace{-3mm}
    \caption{Impact of $\rho$ ($\delta = 0.02,\mu = 1.0$). 
    } 
    \label{fig:impact_rho} 
\end{figure*}

\begin{figure*}[!ht]
    \centering
    \subfigure[Haggle]{
    \includegraphics[width=0.3\textwidth]{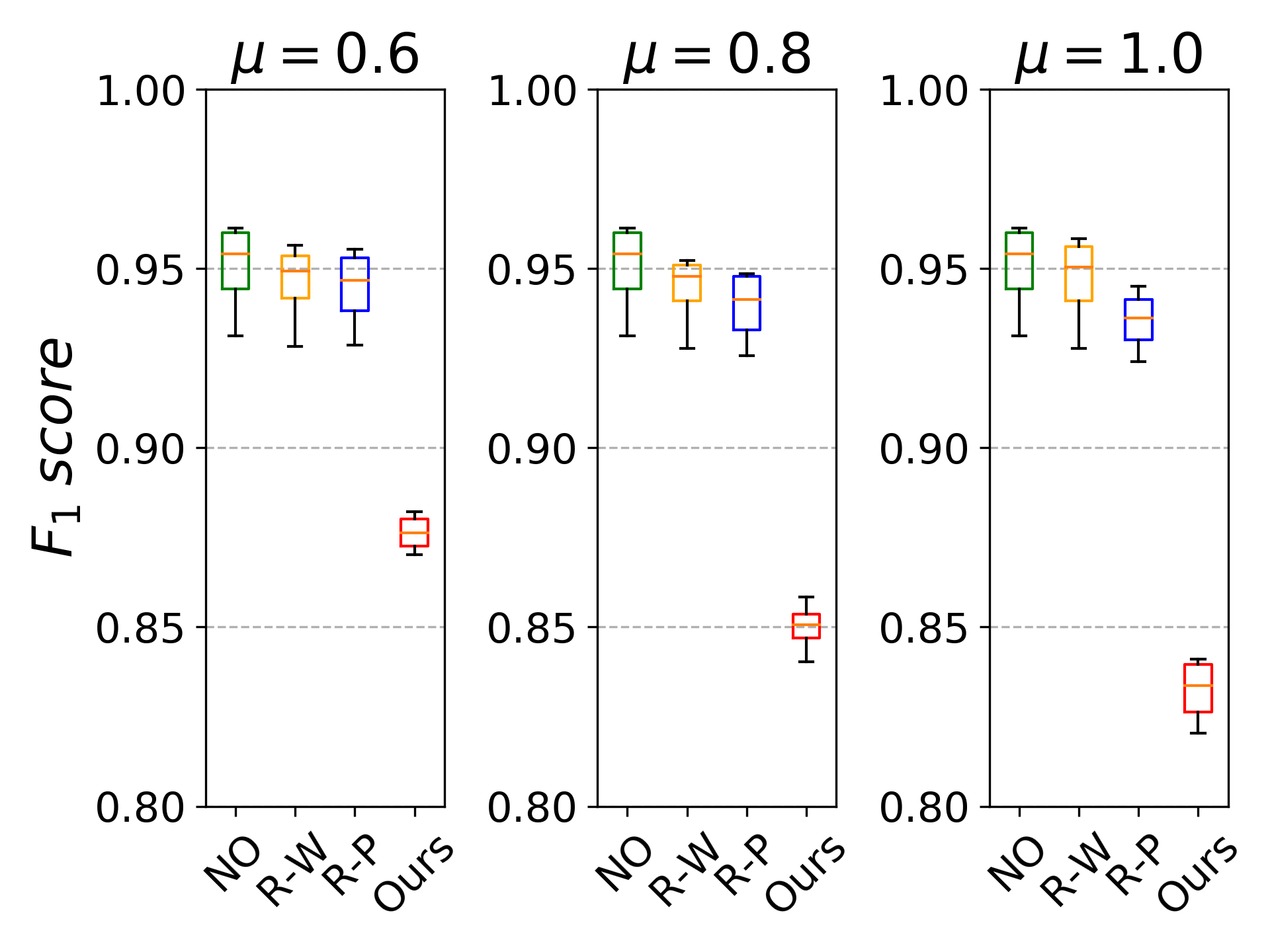}
    }
    \subfigure[Hypertext]{
    \includegraphics[width=0.3\textwidth]{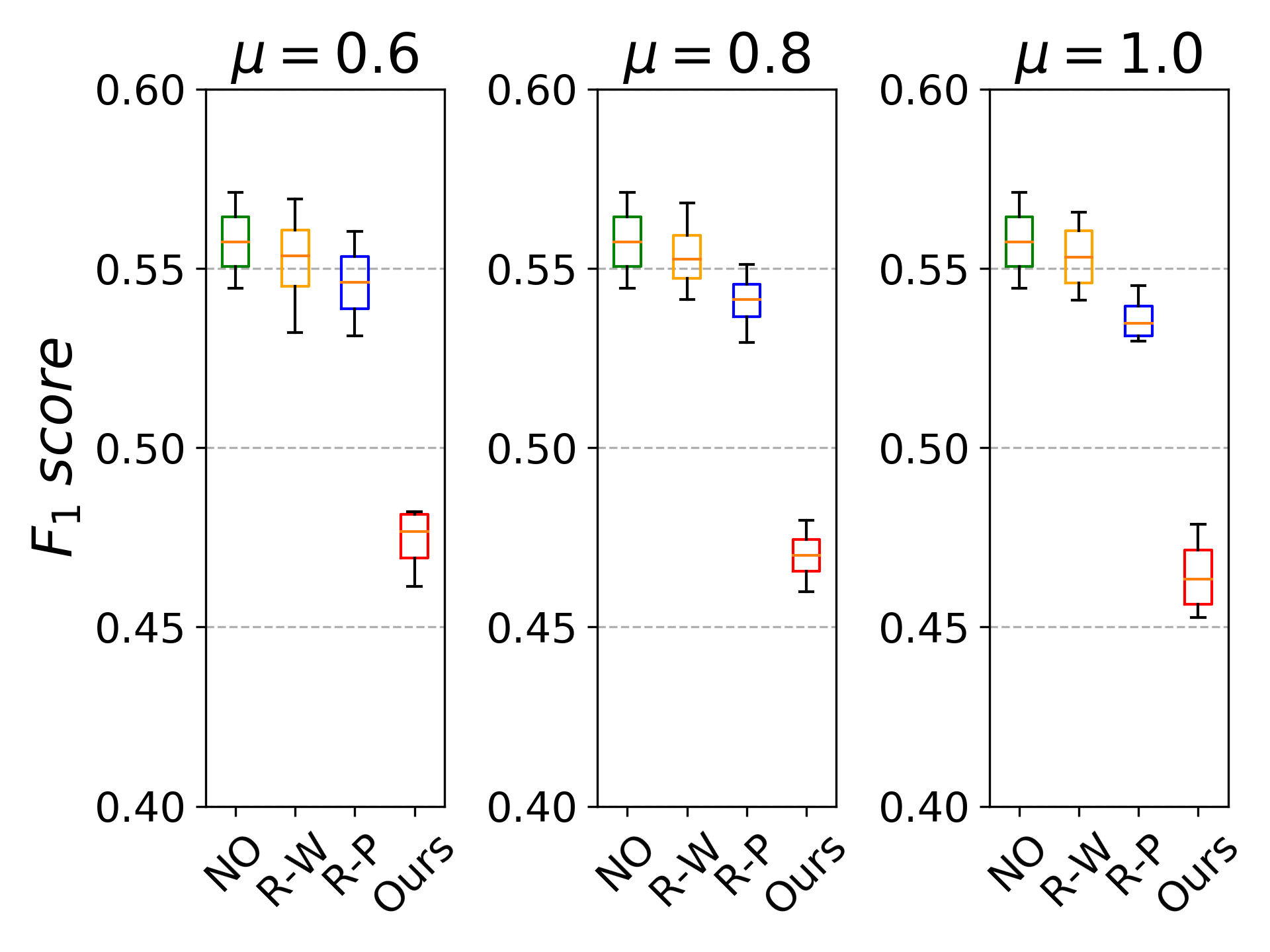}
    }
    \subfigure[Trapping]{
    \includegraphics[width=0.3\textwidth]{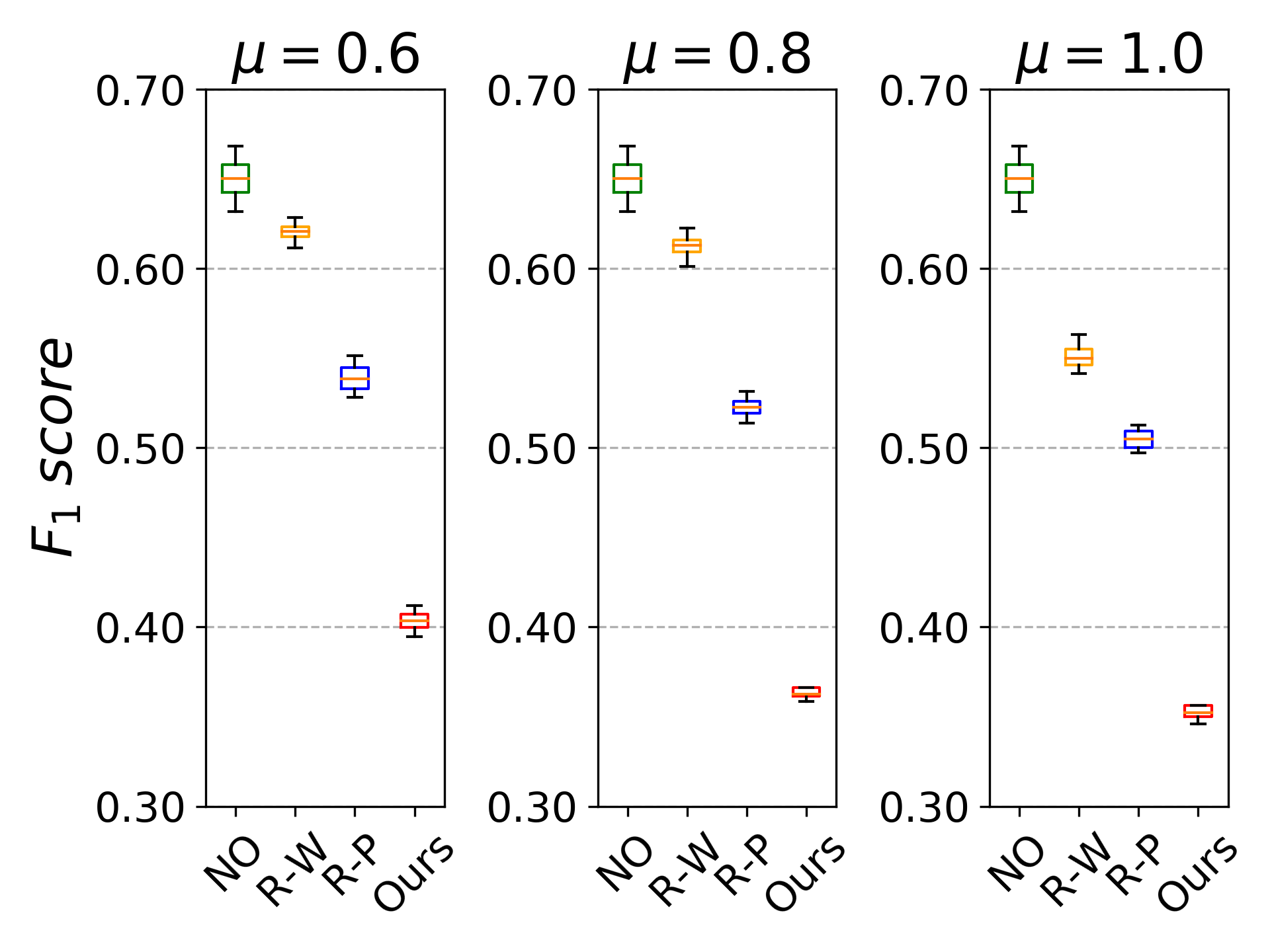}
    }
    \vspace{-3mm}
    \caption{Impact of $\mu$ $(\delta = 0.02,\rho = 0.30)$. 
    } 
    \label{fig:impact_mu}
\end{figure*}

\subsubsection{Configuration.} 
Each dataset is divided into a training set, a validation set, and a testing set, where each training/validation/testing example consists of a sequence of $n=10$ graphs. As each dataset has 30 graphs, we have $20$ examples in total. 
We use $M_{tr}=10$ training examples to train DyGCN; $M_{val}=5$ validation examples to train our attack model; and the remaining $M_{te}=5$ testing examples to evaluate our attack. 
All our experiments are performed on a Linux machine with 128GB memory and 10 cores.

In DyGCN, the number of GCN layers is $L=3$ with each layer having 64 units, and LSTM has 2 layers. As our graph datasets do not have node features, we thus use the identity matrix to represent node feature matrix, similar to~\cite{xu2018powerful}.  
In our attack, there are several key parameters that could affect the attack performance: the fraction $\mu$ of total graphs to be perturbed in an example; the fraction $\rho$ of total nodes to be selected as the popular/neglected nodes; and the fraction $\delta$ of total edges to be perturbed in each graph. 
By default, we set $\mu=1.0$, $\rho=0.30$, and $\delta=0.02$.
We also explore the impact of these parameters. 
We fix the other parameters as the default value when we study one specific parameter.  
Specifically, we set $\mu=\{0.6, 0.8, 1.0\}$, $\rho=\{0.15, 0.30, 0.45\}$, and $\delta = \{0.01, 0.02, 0.05\}$.

\subsubsection{Compared attacks.} There are no existing works on attacking dynamic link prediction in the black-box setting. 
Here, we propose two baseline attacks and compare them with our  attack under the same setting (e.g., the same number of graphs to be attacked, the same attack budget). 

\begin{itemize}
    \item \textbf{Random-whole.} In this attack, the attacker  randomly deletes an edge from and adds an edge to the \emph{whole} graph in each attack step.
    
    \item \textbf{Random-partial.} In this attack, in each attack step, the attacker  randomly selects a pair of nodes from $V_{p}$ and deletes the edge if an edge exists between them; and selects another pair of nodes from $V_{n}$ and adds the edge if there is no edge between them.
    
\end{itemize}

\subsubsection{Evaluation metric.} 
Similar to existing works~\cite{chen2019lstm,pareja2020evolvegcn,manessi2020dynamic}, we use $F_1$ score to evaluate the performance of a link prediction algorithm.  
Thus, the function $f$ in Equation~\ref{obj:reward} is the $F_1$ score.  
The higher/lower the $F_1$ score, the better/worse the link prediction method is. 
All results are reported in an averaged $F_1$ score on the testing examples.   

\begin{table}[tbp]
    \centering
    \caption{\textbf{Link prediction results on clean data}
    \vspace{-3mm}
  }  
    \begin{tabular}{c|ccc}
        \toprule
        {\bf Dataset} & {\bf Haggle} & {\bf Hypertext} & {\bf Trapping} \\
        \midrule
        {\bf No attack(NO)} & $0.9502$ & $0.5576$ & $0.6501$\\
        {\bf Random-whole(R-W)} & $0.9467$ & $0.5532$ & $0.5512$ \\
        {\bf Random-partial(R-P)} & $0.9353$ & $0.5361$ & $0.5024$ \\
        {\bf Our attack(Ours)} & $0.8322$ & $0.4645$ & $0.3542$ \\
        \bottomrule
    \end{tabular}
    \label{tab:comp}
    \vspace{-4mm}
\end{table}

\subsection{Attack Results on DyGCN}

\subsubsection{Effectiveness of our attack.}
Table~\ref{tab:comp} shows DyGCN's performance on the clean graphs, i.e., no attack, and the perturbed graphs, i.e., with our attack and two random attacks.   
We observe that our attack is effective. For instance, compared with no attack, our attack has an $12.4\%$, $16.7\%$, and $45.5\%$ $F_1$ score  degradation on the three datasets, respectively. 
Moreover, our attack is much more effective than random attacks. For instance, on Trapping, our attack has a $30.3\%$ and $22.8\%$ performance gain over Random-whole attack and Random-partial attack, respectively.

\subsubsection{Impact of $\delta$.}
Figure~\ref{fig:impact_delta} shows DyGCN's $F_1$ score under all compared attacks vs. fraction $\delta$ of total edges perturbed.
We observe that as $\delta$ increases, all attacks have better attack performance. This is because, a larger $\delta$ indicates that an attacker is capable of perturbing a larger number of links/edges. Moreover, our attack is much more effective than random attacks at a given $\delta$. For instance, when only $1\%$ edges can be perturbed, our attack can reduce the $F_1$ score with $8.8\%$, $9.0\%$, $37.9\%$, while random attacks reduce around $1.3\%$, $1.4\%$, and $16.6\%$, on the three datasets, respectively. 

\subsubsection{Impact of $\rho$.}
Figure~\ref{fig:impact_rho} shows DyGCN's $F_1$ score under all compared attacks vs. fraction $\rho$ of total nodes selected as the popular/neglected node set.  
Similarly, we observe that as $\rho$ increases, all attacks' performance increases. However, our attack requires a much smaller $\rho$ than random attacks, in order to achieve a promising attack performance.

\subsubsection{Impact of $\mu$.}
Figure~\ref{fig:impact_mu} shows DyGCN's $F_1$ score under all  compared attacks vs. fraction $\mu$ of total graphs perturbed in each example. Similarly, we observe that as $\mu$ increases, $F_1$ score of all attacks decreases. 
However, our attack obtains a good attack performance when perturbing a relatively smaller number of graphs (e.g., $\mu=0.6$).

\begin{figure}[t]
    \centering
    \includegraphics[width=0.4\textwidth]{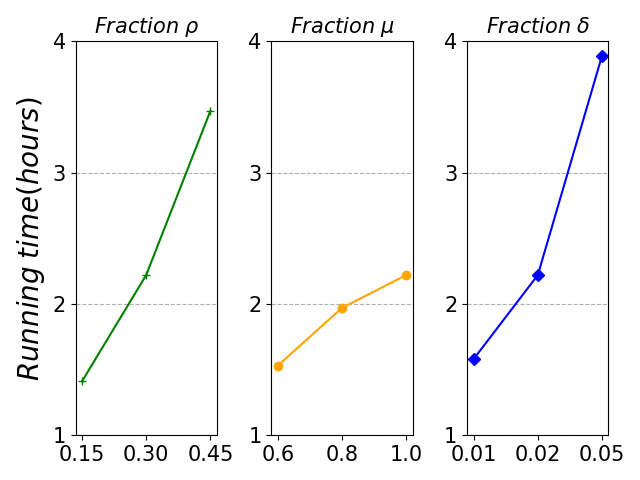}
    \vspace{-3mm}
    \caption{Running time of training our attack vs. $\rho$, $\mu$, and $\delta$. 
    } 
    \label{fig:time} 
\end{figure}

\subsubsection{Efficiency of our attack.} 
We use running time as the metric to evaluate the efficiency of training our attack. 
Figure~\ref{fig:time} shows the running time vs. $\rho$, $\mu$, and $\delta$ on Haggle. When we observe one fraction, the other two are set to default values. Note that we have similar observations on the other two datasets, and thus show results on Haggle for simplicity.  
We observe that the running time is linear to the three parameters. Thus, we can conclude that our attack is also efficient.

\section{Conclusion}
\label{sec:conclusion}

We propose the first black-box evasion attack against graph neural network-based link prediction in dynamic graphs (LPDG). 
Our attack aims to perturb the graph structure so as to fool the LPDG model, while not knowing the model parameters, model  architecture, etc.
We formulate our attack as an optimization problem, which is NP hard, and then develop a reinforcement learning-based method to realize our attack. 
Experimental results on three real-world graph datasets show that our attack is both effective and efficient.

\section{Related Work}
\label{sec:relwork}

We review existing works that perform attacks against graph data. These methods can be summarized as attacking graph-based clustering~\cite{chen2017practical}, graph-based collective classification~\cite{torkamani2013convex,wang2019attacking}, graph embedding~\cite{dai2018adversarialnet,sun2018data,chen2018fast,bojchevski2019adversarial,chang2020restricted}, and graph neural networks~\cite{zugner2018adversarial,dai2018adversarial,zugner2019adversarial,wang2019attacking,wu2019adversarial,xu2019topology,sun2020adversarial,zhang2020backdoor}.   
For instance, \cite{chen2017practical} designed a practical attack against spectral clustering, a type of graph-based clustering methods.   ~\cite{wang2019attacking} proposed an attack against the collective classification method, called linearized belief propagation, by manipulating the graph structure. 
Z{\"u}gner et al.~\cite{zugner2018adversarial} developed a Nettack method to attack graph convolutional network (GCN)~\cite{kipf2016semi} by perturbing both the node features and graph structure.
However, we note that all these attacks focus on attacking a single static graph. 

To the best of our knowledge, only one work~\cite{chen2019time} studies  adversarial attacks to link prediction in dynamic graphs. However, this attack is white-box (i.e., the attacker knows all information about the trained model), and is specially designed for a specific LPDG method called deep dynamic graph embedding. In contrast, our attack is black-box and is applicable to arbitrary LPDG methods. 
In our work, for simplicity, we focus on attacking the state-of-the-art DyGCN method~\cite{manessi2020dynamic}.

\bibliographystyle{plain}
\bibliography{refs}

\end{document}